\documentclass[]{aipproc}
\def\smallfrac#1#2{\hbox{${{#1}\over {#2}}$}}
\def\as{\alpha_s}
\def\ah{\widehat\alpha_s}
\def\bea{\begin{eqnarray}}
\def\eea{\end{eqnarray}}

\def\half{\hbox{${1\over 2}$}}

\catcode`@=11 %This allows us to modify plain macros

\def\tozero#1{\mathrel{\mathop{\sim}\limits_{\scriptscriptstyle {#1\rightarrow0 }}}}
\def\slash#1{\mathord{\mathpalette\c@ncel#1}}
 \def\c@ncel#1#2{\ooalign{$\hfil#1\mkern1mu/\hfil$\crcr$#1#2$}}
\def\lsim{\mathrel{\mathpalette\@versim<}}
\def\gsim{\mathrel{\mathpalette\@versim>}}
 \def\@versim#1#2{\lower0.2ex\vbox{\baselineskip\z@skip\lineskip\z@skip
       \lineskiplimit\z@\ialign{$\m@th#1\hfil##$\crcr#2\crcr\sim\crcr}}}
\catcode`@=12 %at signs are no longer letters

\SetInternalRegister\hbadness{8000} % pseudo latin isn't breaking very well :-)

\newcommand\doingARLO[2][]{%
  \ifx\mmref\undefined #1\else #2\fi
}
  
\layoutstyle{6x9}
\begin{document}
\hoffset-1.7cm

\setcounter{page}{0}
\begin{flushright}
{\tt hep-ph/0109235}\\
{RM3-TH 01-12}\\Edinburgh 2001-14\\
\end{flushright}
%\vglue.3cm
\title {Theoretical Aspects of HERA Physics}

\author{Stefano Forte\thanks{On leave from INFN, Sezione di Torino, Italy}~~}{
  address={INFN, Sezione di Roma III, via della Vasca Navale 84,
  I--00146 Roma, Italy} 
}
\iftrue
\author{Richard D.~Ball}{
  address={Department of Physics and Astronomy,
University of Edinburgh, Edinburgh EH9 3JZ, Scotland}}
\fi

\copyrightyear  {2001}

%\vglue.7cm

\begin{abstract}
We discuss the theoretical underpinnings for the extraordinary success of perturbative QCD in the
description of HERA data. In particular, we examine recent progress in the understanding of
perturbative QCD at small $x$. We explain the relation between evolution equations in $Q^2$ and $x$, and how
they can be used for simultaneous resummation of the relevant large logs at HERA. We show that while
the HERA data can be understood within our current knowledge of the perturbative expansion of
QCD, they pose stringent constraints on the perturbatively inaccessible behaviour of QCD in the
Regge limit. 
\end{abstract}

\date{\today}

\maketitle
\vglue3.cm
\begin{center}
Presented at the workshop
{\bf QCD@WORK}\\
Martina Franca, Italy, June 2001\\
\smallskip
{\it to be published in the proceedings}
\end{center}
\addtocounter{footnote}{-1}

\newpage 
\maketitle
\section{Perturbative QCD at HERA}

QCD has been tested at HERA~\cite{threv,exprev} 
over the last several years to an accuracy which is now 
comparable to that of tests of the electroweak sector at LEP: 
perturbative QCD turns out to provide an embarrassingly
successful description of the HERA data, even in kinematic regions
where 
simple fixed--order perturbative predictions should
fail. This success is most strikingly demonstrated by the comparison
with the data of the scaling violations of structure functions
predicted by the QCD evolution equations~\cite{zeus,hone}: %(see Fig.~1): 
the data agree
with the theory over five orders of magnitude in both $x$ and $Q^2$.
%\begin{figure}
%\includegraphics[width=0.45\linewidth,clip]{h1f2.ps} 
%\includegraphics[width=0.45\linewidth,clip]{zeusf2.ps} 
%\caption{Comparison between the predicted scaling violations and the
%data of the H1~\cite{hone} (left) and ZEUS~\cite{zeus} (right)
%experiments, together with older fixed--target data.}
%\end{figure}

The significance of this sort of result is somewhat obscured by the need to fit
the shape of parton distributions at a reference scale, which might suggest
that deviations from the predicted behaviour could be
accommodated by changing the shape of the parton
distribution. However, this is not true because of the predictive
nature of the QCD result: given the shape of  partons at one scale,
there is no freedom left to fit the data at other scales. This
predictivity is particularly transparent in the small $x$ region, where the
fixed--order QCD result actually becomes asymptotically
independent of the parton distribution, apart  from an overall
normalization. Indeed, the data for $\ln F_2$
plotted versus the variable
$\sigma\equiv\ln{x_0\over x}\ln {\as(Q^2_0)\over \as(Q^2)}$ are
predicted to lie on a straight line, with universal slope 
$2\gamma=12/\sqrt{33-2 n_f}$ (double asymptotic
scaling~\cite{das,DGPTWZ}). 
The
predicted scaling is spectacularly borne out by the data, as shown in
fig.~1: in fact, the data are now so accurate that one can see the change
in slope when passing the $b$ threshold, and indeed double scaling is
only manifest if one separates data in the regions where $\alpha_s$ runs
with  $N_f=4$ from those with $N_f=5$.\footnote{The fact that the
observed slope is somewhat smaller than the predicted one, especially
at low $Q^2$, is due to NLO corrections~\cite{zako} as
well as corrections due to the ``small'' eigenvalue of perturbative
evolution~\cite{lech}.}  Equally good agreement with fixed--order
perturbation theory is seen when considering less inclusive observables.
\begin{figure}
%\begin{center}
\includegraphics[width=0.48\linewidth,clip]{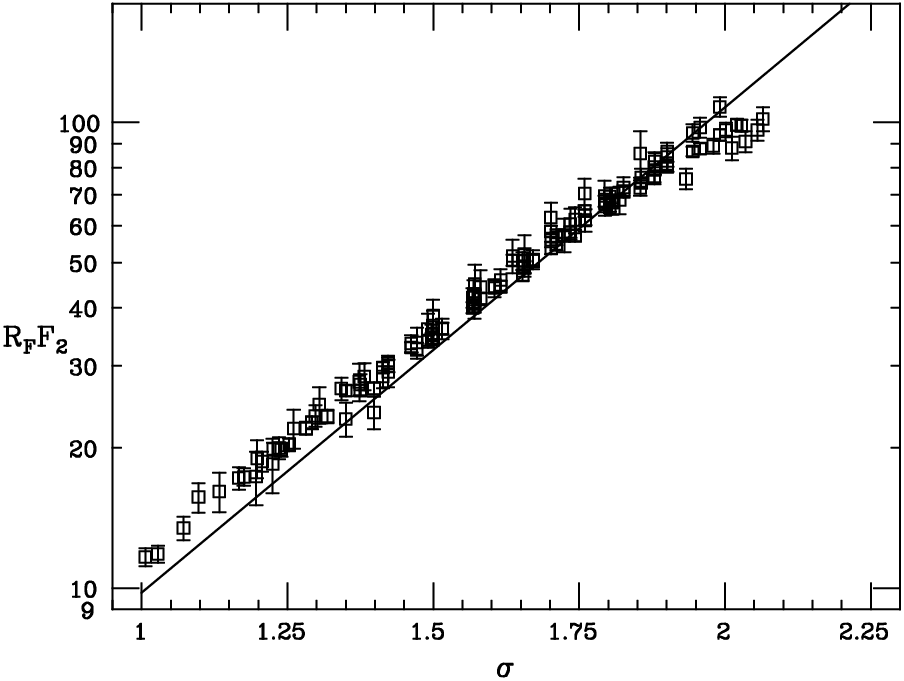} 
\includegraphics[width=0.45\linewidth,clip]{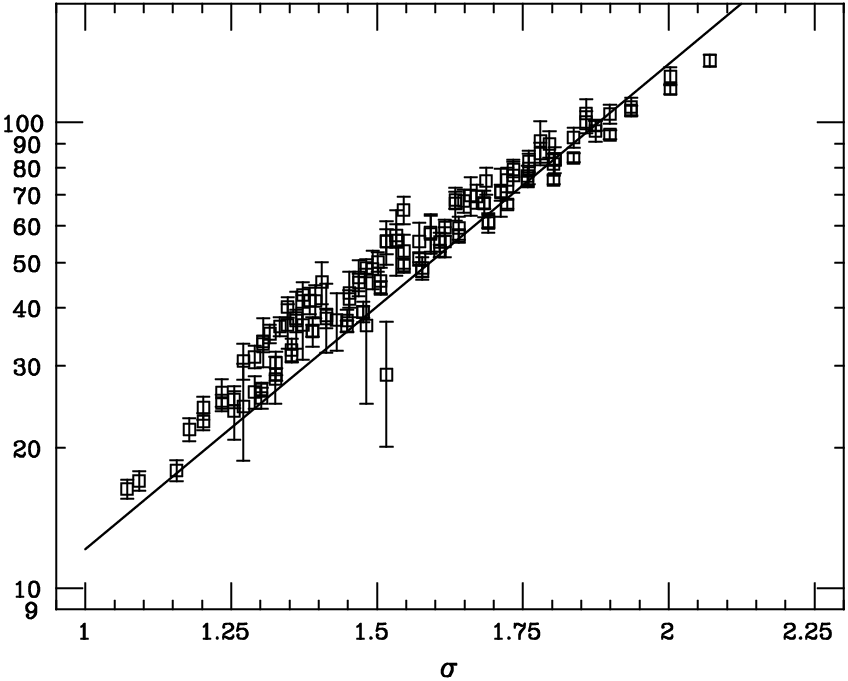} 
\caption{Double asymptotic scaling of the H1 data~\cite{hone}. The
scaling variable $\sigma\equiv\ln(x_0/ x)\ln (\as(Q_0^2)/\as(Q^2))$
is defined with $x_0=0.1$, $Q_0=1~{\rm GeV}$; the rescaling factor
$R_F$ is as in Ref.~\cite{das}. Only data with
$\rho\ge 1$, $\sigma\ge 1$, $Q^2\ge4~{\rm GeV}^2$; $x\le0.03$ are
plotted. Left: $Q^2\le m_b^2$; right: $Q^2> m_b^2$. The straight line
is the asymptotic prediction.}
%\end{center
\end{figure}

This agreement of the data with fixed--order perturbative QCD
computations is very surprising, in that the perturbative expansion
receives contributions of order $\as\ln{1\over x}$ so one would expect
higher--order corrections to be non--negligible whenever
$\as\ln{1\over x}\gsim 1$, i.e. in most of the HERA region. As is well
known, the resummation of leading $\ln{1\over x}$ (LLx) contributions
to gluon--gluon scattering, and thus to a wide class of hard
processes, including small $x$ scaling violations of structure
functions, is accomplished by means of the BFKL evolution
equation~\cite{bfkla,bfklb,bfklc}. Matching the BFKL approach to
standard perturbative computation, however, is
nontrivial~\cite{summing,EHW}, while the BFKL equation itself
seems to be unstable towards the inclusion of higher order
corrections~\cite{fl}. Hence, the main problem in understanding HERA
physics, i.e. perturbative QCD at small $x$ is that of establishing
``consistency of the BFKL approach with the more standard DGLAP~\cite{ap,gl}
evolution equations''~\cite{lmcl}, which embody
the leading $\ln Q^2$ (LLQ$^2$) resummation on which perturbative QCD
is based. 
This problem is now solved~\cite{sxap,sxres,ccs}, and on
the basis of this solution it is possible to combine the available
information on perturbation theory at small $x$, and use it to explain
the unexpected success of fixed--order calculations.

\section{Duality}

Let us for definiteness consider the prototype problem of the
description of small
$x$ scaling violations of parton distributions. For simplicity,
consider the case of a single parton distribution $G(x,Q^2)$, which can
be thought of as the dominant eigenvector of
perturbative evolution. Scaling violations are then described by the
Altarelli-Parisi equation satisfied by $G(x,Q^2)$, and thus summarized
by the Altarelli--Parisi splitting function $P(x,\as)$~\cite{ap}.

The basic result which allows the study of scaling violations at small
$x$ is {\it duality} of perturbative
evolution~\cite{afp,sxres,sxphen}, 
namely, the fact 
that, because the Altarelli-Parisi equation 
is an integro--differential
equation in the two variables $t\equiv\ln Q^2/\Lambda^2$ and
$\xi\equiv 1/x$,  it can be equivalently 
cast in the form of a differential equation in $t$
satisfied by the $x$--Mellin transform 
\begin{equation}
G(N,t)=\int^{\infty}_{0}\! d\xi\, e^{-N\xi}~G(\xi,t),
\label{nmel}
\end{equation}
or a differential equation in $\xi$
satisfied by the $Q^2$--Mellin transform 
\begin{equation} 
G(\xi,M)=\int^{\infty}_{-\infty}\! dt\, e^{-Mt}~G(\xi,t)
\label{mmel} 
\end{equation} 
of the parton distribution.
The pair of dual evolution equations are 
\begin{eqnarray}
\frac {d}{dt}G(N,t)&=&\gamma(N,\as)~G(N,t)
\label{tevol}\\
\frac {d}{d\xi}G(\xi,M)&=&\chi(M,\as)~G(\xi,M),
\label{xevol}
\end{eqnarray}
where eq.~(\ref{tevol}) is the standard renormalization--group equation,
with anomalous dimension $\gamma(N,t)$, and eq.~(\ref{xevol}) is
essentially the BFKL equation. Duality is the statement that 
the solutions of these two equations coincide to all perturbative
orders,  up to power suppressed corrections,  provided their
kernels are related by 
\begin{equation}
\chi(\gamma(N,\as),\as)=N.\label{dual}
\end{equation}
This means that the BFKL and Altarelli-Parisi equations 
describe the same physics: it is the
choice of the kernel to be used in the evolution equation which
determines which is the large scale which is resummed. 
We can then discuss the construction and resummation of the kernel 
irrespective of the specific evolution equation where it is used, with
the understanding that the kernel can be equivalently viewed as a
$\gamma(N,\as)$ or a $\chi(M,\as)$, the two being related by
eq.~(\ref{dual}). Before doing this, we sketch how duality can be proven
order by order in perturbation theory.

\subsection{Fixed coupling}

Perturbative duality is most easy to prove when the coupling does not run,
since in this case the two scales $t$ and $\xi$ appear in the
Altarelli--Parisi equation 
in a completely symmetric way. It is  convenient to introduce the
double--Mellin transform $G(N,M)$ of the parton distribution. The
solution to the Altarelli--Parisi equation in $M,N$ space has the form
(which can be e.g. obtained by performing an $M$--mellin transform
eq.~(\ref{mmel}) of the solution to the renormalization--group
eq.~(\ref{tevol}))
\begin{equation}
G(N,M)=\frac{G_0(N)}{M-\gamma(N,\as)},
\label{tsoln}
\end{equation}
where $G_0(N)$ is a boundary condition at a reference scale
$\mu^2$. 

The inverse Mellin transform of eq.~(\ref{tsoln}) coincides
with the residue of the simple pole in the $M$ plane of $e^{t M} G(N,M)$,
and thus its
scale dependence 
is entirely determined by the location of the simple pole of
$G(N,M)$~(\ref{tsoln}) , namely, 
the solution to the equation
\begin{equation}
M=\gamma(N,\as). 
\label{pole}
\end{equation} 
The pole condition Eq.~(\ref{pole}) can be equivalently viewed
as an implicit equation for $N$: $N=\chi(M,\as)$, where $\chi$ is
related to $\gamma$ by eq.~(\ref{dual}). Hence, the function
\begin{equation}
G(N,M)=\frac{F_0(N)}{N-\chi(M,\as)},
\label{xisoln}
\end{equation}
corresponds to the same
$G(t,x)$ as eq.(\ref{tsoln}), because the location of the  respective poles 
in the $M$ plane are the same, while the residues are also the same,
provided the
boundary conditions are matched by 
\begin{equation}
 G_0(N)=-
{ F_0(\gamma(\as,N))\over\chi^\prime(\gamma(\as,N))}.
\label{bcmatch}\end{equation}

Eq.~(\ref{xisoln}) is immediately recognized as the $N$-Mellin
of the solution to the evolution equation~(\ref{xisoln}) with
boundary condition $F_0(M)$
(at some reference $x=x_0$), which is what we set out to prove. 
In general, the analytic continuation of the function $\chi$ defined by
eq.~(\ref{dual}) will be such that eq.~(\ref{pole}) has more than one
solution (i.e. $\gamma$ is multivalued). In this case, poles further
to the left in the $M$ plane correspond to power--suppressed
contributions, while poles to the right correspond to contributions
beyond perturbation theory (they do not contribute when the inverse
$M$--Mellin integral is computed along the integration path which
corresponds to the perturbative region).

It is easy to see that upon duality the leading--order
$\chi=\as\chi_0$ is mapped onto the leading singular
$\gamma=\gamma_s(\as/N)$, and  conversely the leading--order 
$\gamma=\as\gamma_0$ is mapped onto the leading singular
$\chi=\chi_s(\as/M)$. In  general, the expansion of $\chi$ in
powers of $\as$ at fixed $M$ is mapped onto
the expansion of $\gamma$ in
powers of $\as$ at fixed $\as/N$, and conversely. 
So in particular at LLQ$^2$ it is
enough to consider $\gamma_0$ or $\chi_s$, and at
LLx it is enough to consider $\gamma_s$ or $\chi_0$. The running of
the coupling is a LLQ$^2$ but NLLx effect, so beyond LLx the
discussion given so far is insufficient.

\subsection{Running coupling}

The generalization of duality to the running coupling case is
nontrivial because the running of the coupling breaks the symmetry of
the two scales $\xi$ and $t$ in the Altarelli--Parisi
equation. Indeed, upon $M$--Mellin transform~(\ref{mmel}) the usual
one--loop running
coupling becomes the differential operator
\begin{equation}
\ah = \frac{\as}{1-\beta_0 \as \smallfrac{d}{dM}}+\cdots,
\label{ahdef}
\end{equation}
where $d\as/dt= -\beta_0\as^2$. 

Consider for simplicity the LLx
$x$--evolution equation, i.e. eq.~(\ref{xevol}) with
$\chi=\as\chi_0(M)$, and include running coupling effects by replacing
$\as$ with the differential operator eq.~(\ref{ahdef}). We can 
solve the equation perturbatively by expanding the solution  
in powers of $\as$ at
fixed $\as/N$: the leading--order  solution is given by
eq.~(\ref{xisoln}), the next--to--leading order  is obtained by
substituting this back into the equation and retaining terms up to
order $\beta_0\as$, and so on~\cite{sxap}. We can then determine the associate
$G(N,t)$ by inverting the $M$--Mellin, and try to see whether this
$G(N,t)$ could be obtained as the solution of a renormalization group
(RG) equation~(\ref{tevol}). 

The inverse Mellin is again given by the residue of the pole of 
$e^{t M} G(N,M)$ in the M--plane, where $G(N,M)$ is now the
perturbative solution. When trying to  identify this
with a solution to eq.~(\ref{tevol}) there are two potential sources
of trouble. The first is that now the
perturbative solution at order $(\as\beta_0)^n$
has a $(2n+1)$--st order pole. Therefore,  the scale--dependence of the inverse
Mellin is now a function of both $\as$ and $t$, whereas
the solution of a RG equation depends on $t$ only through the running
of $\as$. Hence it is not obvious that a dual anomalous dimension
will exist at all. The second is
that even if a dual $\gamma$ does exist, it is not obvious that it will
depend only on $\chi$ and not also on the boundary condition
$F_0(M)$ eq.~(\ref{xisoln}): in such case, the running of the coupling
in the $\xi$--evolution equation
would entail a breaking of factorization.

However, explicit calculation shows that it is possible to match
the anomalous dimension and
the boundary condition order by order in perturbation
theory in such a way that
both duality and
factorization are respected. Namely,
the solution to the leading--twist running coupling
$x$--evolution eq.~(\ref{xevol}) with kernel $\ah \chi_0$ and
boundary condition $G_0(M)$ is the same as that of the
renormalization group eq.~(\ref{tevol}) with boundary conditions and
anomalous dimension given by
\begin{eqnarray}
&&\gamma(\as(t),\as(t)/N)=\gamma_s(\as(t)/N)+\as(t)\beta_0\Delta\gamma_{ss}(\as(t)/N)+\nonumber\\
&&\quad\qquad+
(\as(t)\beta_0)^2 \Delta\gamma_{sss}(\as(t)/N) +O(\as(t)\beta_0)^3\\
\label{efgam}
&&G_0(\as,N)=G_0(N)+\as\beta_0 \Delta^{(1)}G_0(N)+
(\as\beta_0)^2\Delta^{(2)} G_0(N)
+O(\as\beta_0)^3,
\label{efbc}
\end{eqnarray}
where the leading terms $\gamma_s$ and $G_0(N)$ are given
by eqs.~(\ref{dual}) and (\ref{bcmatch})
respectively. The subleading corrections are
\begin{eqnarray}
\Delta\gamma_{ss}&=&-
\frac{\chi_0''\chi_0}
{2\chi_0'^2}\\
\Delta^{(1)}G_0(N)&=&{2{\chi_0'}^2
F_0-\chi_0\left(F_0'\chi_0''-\chi_0'F_0''\right)\over 2{\chi_0'}^3},
\label{nlocorr}
\end{eqnarray}
where all derivatives are with respect to the arguments of $\chi_0(M)$
and $F_0(M)$, which are then evaluated as functions of $\gamma_s(\as/N
)$. 
The sub--subleading correction to the anomalous dimension is
\begin{equation}
\Delta\gamma_{sss}=-\chi_0^2{15{\chi_0''}^3-16
\chi_0'\chi_0''\chi_0'''+3 {\chi_0'}^2\chi_0''''\over 24 {\chi_0'}^5},
\label{nnlocorr}
\end{equation}
and we omit the very lengthy expression for $\Delta^{(2)}G_0(N)$.
The fact that duality and factorization hold up to NNLLx is
nontrivial, and suggests that they should hold to all orders. An
all--order proof can be in fact constructed~\cite{sxrun}.

Once the corrections to duality eq.~(\ref{efgam}) are determined, they
can be formally re-interpreted as additional contributions to $\chi$:
namely, one can impose that the duality eq.~(\ref{dual}) be
respected, in which case the kernel to be used in it is an
``effective'' $\chi$,
obtained from the kernel of the $x$--evolution eq.~(\ref{xevol}) by
adding to it running coupling corrections order by order in
perturbation theory:
$\chi_0$ will be free of such correction, $\chi_1$ will receive a
correction 
\begin{equation}
\Delta\chi_1=\beta_0{1\over
2}{\chi_0(M)\chi_0''(M)\over{\chi_0'}^2(M)},
\label{efchi}
\end{equation}
and so forth. Applying duality to
the known  one--loop anomalous dimensions $\gamma_0$ thus
gives us the resummation of the all--order singular contributions
$\chi(\as/M)$ to this effective $\chi$, which include the running
coupling correction eq.~(\ref{efgam}) and its higher--order generalizations.

\section{Resummation}
Because the first two orders of the expansion of $\chi$ in
powers of $\as$ at fixed $M$ and of the expansion of $\gamma$ in
powers of $\as$ at fixed $N$ are known, it is possible to exploit duality
of perturbative evolution to combine this information 
into anomalous dimension which accomplish the simultaneous resummation of 
leading and next--to--leading logs of $x$ and $Q^2$. 
In fact, it turns out that both a small $M$ and a small $N$
resummation of anomalous dimensions are necessary in order to obtain a
stable perturbative expansion, while unresummed anomalous dimensions
leads to instabilities. Both sources of instability are generic
consequences of the structure of the perturbative expansion, and 
could have been predicted before the actual explicit computation~\cite{fl} of
subleading small-$x$ corrections.

\subsection{Small M}
The perturbative expansion of $\chi$ at fixed $M$ is very badly
behaved in the vicinity of $M\sim0$: at $M=0$,
$\chi_0$ has a simple pole, $\chi_1$ has a double pole and so on.
In practice, this spoils the
behaviour of $\chi$ in most of the physical region $0< M<
1$. Because $1/M^k$ is the Mellin transform of ${\Lambda^2\over Q^2}
{1\over k!}\ln^{k-1}(Q^2/\Lambda^2)$, these singularities correspond to
logs of $Q^2$ which are left unresummed in a LLx or NLLx
approach~\cite{salam}.  

The resummation of these contributions may be understood in terms of
momentum conservation, which implies that $\gamma(1,\as)=0$ (note our
definition of the $N$--Mellin transform~(\ref{nmel}), and also that
$\gamma$ is to be identified with the large eigenvector of the
anomalous dimension matrix).  The duality eq.~(\ref{dual}) then
implies that a momentum--conserving $\chi$ must satisfy
$\chi(0,\as)=1$. This, together with the requirement that $\chi$
admits a perturbative expansion in  powers of $\as$, implies that in
the vicinity of $M=0$, the generic behaviour of the kernel is
\begin{equation}
\chi_s\tozero M {\as\over\as+\kappa M}={\as\over\kappa M}-{\as^2\over
(\kappa M)^2}+{\as^3\over(\kappa M)^3}+\dots\quad,
\label{momcons}
\end{equation} 
where $\kappa$ is a numerical constant which turns out to be
$\kappa=\pi/C_A$. 
Hence we understand that there must be an alternating--sign 
series of poles at $M=0$, which
sums up to a regular behaviour. 
In fact, we can systematically resum singular contributions to $\chi$
to all orders in $\as$ by including in $\chi$ the terms
$\chi_s(\as/M)$ derived from the leading order $\gamma_0(N)$, and
similarly at next--to--leading order, and so on. Because
the usual anomalous dimension automatically respects
momentum conservation order by order in $\as$,  in order to remove
the small $M$ instability of the expansion of $\chi$ at fixed $M$, it
is sufficient to improve the expansion by promoting it to a ``double
leading'' expansion which combines the expansions in powers of
$\as$ at fixed $M$ and at fixed $\as/M$~\cite{sxres}. 
For example, at
leading order
$\chi=\as\chi_0(M)+\chi_s(\as/M)-{\rm d.c.}$, where the subtraction
refers to the double--counting of the $\as/M$ term which is present
both in $\as\chi_0$ and in $\chi_s(\as/M)$. This expansion of $\chi$
is  dual eq.~(\ref{dual}) to an analogous expansion of
$\gamma$, where at leading order
$\gamma=\as\gamma_0(M)+\gamma_s(\as/M)-{\rm d.c.}$, and so forth. Both
expansions are  well behaved at small $M$, i.e. large $N$.
At this level, it is already clear that the impact of the inclusion of
small-$x$ corrections is moderate: indeed, it turns out that the
double--leading kernel is quite close to the usual two--loop kernel,
except at the smallest values of $N$, i.e. in the neighbourhood of the
minimum of $\chi(M)$~\cite{sxres}.

\subsection{Small N}

The improved double--leading expansion of the anomalous dimension
still requires resummation at small $N$. This is because,
even though the next--to--leading correction to the double--leading 
evolution kernel is small for all fixed $M$, 
it is actually large if $N$ is fixed and small.
This in turn follow from the fact that the leading $\chi$ kernel has a
minimum, so the small $N=\chi$ region corresponds by duality
eq.~(\ref{dual}) to  the vicinity of the minimum where the kernel is
almost parallel to the $\gamma=M$ axis. 

At small $N$, unlike at small $M$, there is no
principle like momentum conservation  which may
 provide a fixed
point of the expansion and thus fix the all--order behaviour. 
The
only way out is thus to treat this all--order behaviour as a free
parameter. Namely, we introduce a parameter $\lambda$ which is equal
to the value of the all--order kernel $\chi$ at its minimum, and then
we expand about this all--order minimum. In practice, this means that
we reorganize the expansion of $\chi$ according to~\cite{sxap} 
\begin{eqnarray}
\chi(M,\as)&=&\as \chi_0(M)+\as^2\chi_1(M)+\dots\nonumber\\
&=&\as \tilde \chi_0(M) +\as^2\tilde\chi_1(M)+\dots,
\label{clamsub}
\end{eqnarray}
where
\begin{equation}
\as\tilde \chi_0(M,\as)\equiv \alpha_s
\chi_0(M)+\sum_{n=1}^\infty \alpha_s^{n+1} c_n,\qquad 
\tilde\chi_i(M)\equiv\chi_i(M)-c_i,
\label{ctil}
\end{equation}
and the constants $c_i$ are chosen in such a way that 
\begin{equation}
\lambda \equiv\as\tilde\chi_0(\half)=\as\chi_0(\half)+\Delta \lambda.
\label{lamdef}
\end{equation}
is the all--order minimum of $\chi$.
Of course, in practice phenomenological predictions will only be
sensitive to the value of $\lambda$ in the region where very small
values of $N$ are probed, i.e. at very small $x$.

\section{Phenomenology}

\begin{figure}
%\begin{center}
\includegraphics[width=0.4\linewidth,clip]{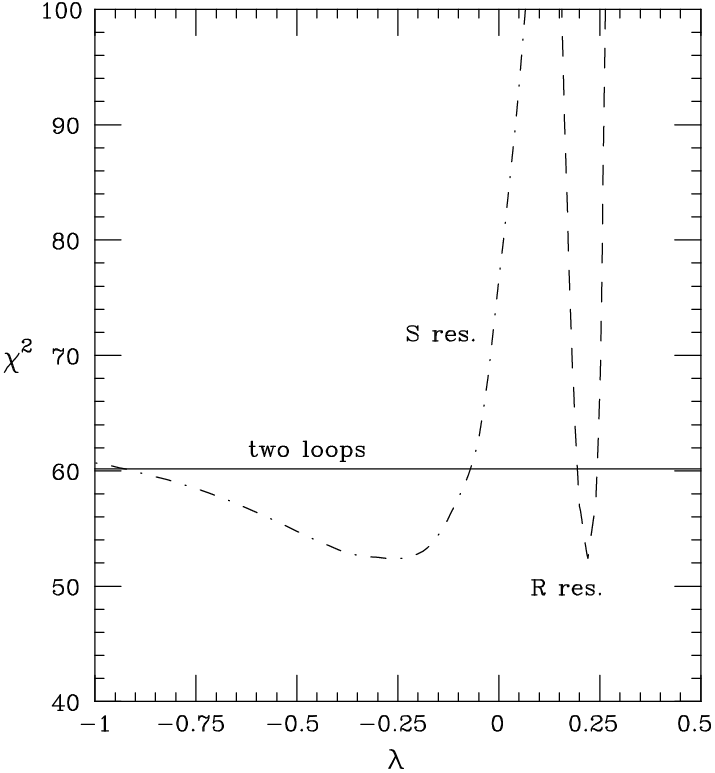} 
\includegraphics[width=0.4\linewidth,clip]{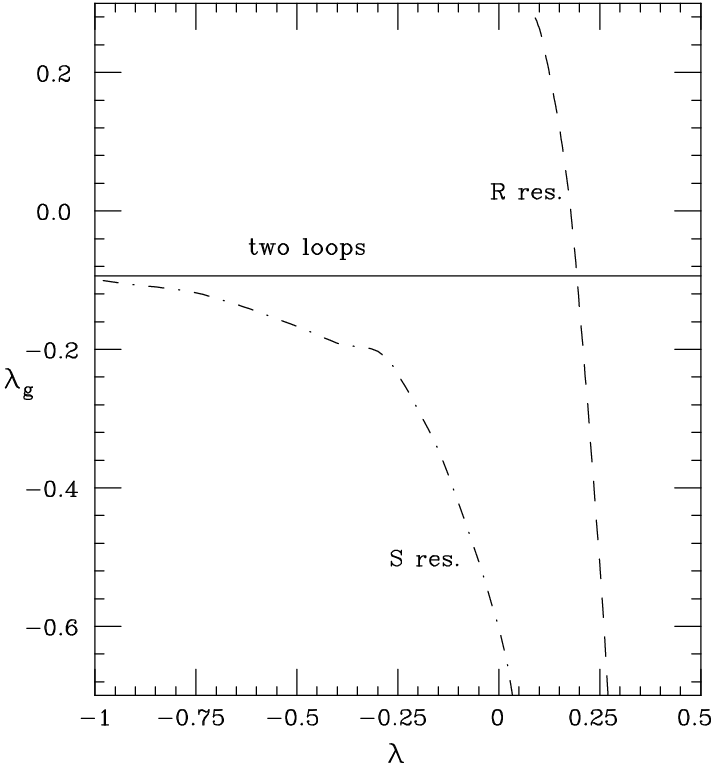} 
\caption{$\chi^2$ (left) and starting gluon slope $G(x, \hbox{4 GeV}^2)\sim
x^{-\lambda_g}$  (right) for the fit~\cite{sxphen}
 to the 95 H1 data~\cite{hone} as a
function of the resummation parameter $\lambda$ eq.~(\ref{lamdef}), for
the two resummation prescriptions discussed in text. The fits are
performed with $\as(M_z)=0.119$.}
%\end{center
\end{figure}

Using duality and the resummation discussed above, one can construct
resummed expressions for anomalous dimensions and coefficient
functions, and wind up with  resummed expressions for physical
observables which may be directly compared to the data.
The need to resum the small $N$ behaviour  entails
that phenomenological predictions will necessarily depend on the
parameter $\lambda$ eq.~(\ref{lamdef}). When the 
resummed double--leading expansion is constructed, a further ambiguity arises
in the treatment of double--counting terms. This
ambiguity is related to the nature of the small
$N$ singularities of the anomalous dimension, which control the
asymptotic small $x$ behaviour.
Specifically, according to the way the double--counting is treated,
the $N=0$ poles of the one-- and two--loop result may survive
in the resummed result (`S--resummation') or not (`R--resummation'). 
Both alternatives
are compatible with the known low--order information on the
evolution kernel, and can be taken as two extreme resummation schemes
which parametrize our ignorance of higher order perturbative terms.
Since the resummed terms also have a  cut
starting at
$N=\lambda$, 
whether or not these low--$N$ poles are present only makes a difference
if $\lambda$ turns out to be small, $\lambda\lsim0.3$.

The $\chi^2$ and starting gluon slope for a fit~\cite{sxphen}  to the
recent H1 data~\cite{hone} 
for the deep--inelastic  cross section are
shown in figure~2, as a function of $\lambda$ and for the two different
resummation prescriptions. It is clear that if the perturbative
$N=0$ poles do not survive the resummation (R resummation) then only a
fine--tuned value of $\lambda\approx 0.2$ is acceptable, whereas if they do
survive (S resummation) essentially any $\lambda\lsim 0$ gives a good
fit. 

Figure 2 demonstrates that it is possible to accommodate the success of
simple  fixed--order approach within a fully resummed scheme, and in
fact the resummed calculation is in somewhat better agreement with the
data than the fixed order one. Even though the effects of the
resummation are necessarily small (otherwise the success of the fixed
order prediction could not be explained) they do have a significant
impact in  the extraction of the parton distribution: the gluon comes
out to be significantly more valence--like than in an unresummed fit.
Hence, the use of resummed perturbation theory is crucial for the
extraction  of reliable parton distributions at small $x$. 

From a theoretical point of view, we see that current data
already pose
 very
stringent constraints on the unknown high--orders of the
perturbative expansion: only a rather soft high--energy
behaviour of the deep-inelastic cross--section is compatible with the
data. Further progress in the understanding of the Regge limit is
likely to require either genuinely nonperturbative input, or an
extension of the standard perturbative domain~\cite{sxap}.

%\subsection{A B-head}
%\footnote{Et iam nox umida caelo praecipitat, suadentque cadentia
%sidera somnos. Et iam nox umida caelo praecipitat, suadentque cadentia
%sidera somnos.} 
%\subsubsection{A C-head}
%\begin{equation}
%J_{ion}=A\frac{exp\left[-\frac{E_a}{kT}\right]}{kT}j \label{ionflux}
%\end{equation}
%\eqref{ionflux}.
%\paragraph{And a D-head}
%\cite{Brown2000,BrownAustin:2000}. Quis talia fando

\begin{theacknowledgments}
A sizable part of this paper is based on  work done in collaboration with
G.~Altarelli. S.F. thanks G.~Nardulli 
for organizing a
very stimulating workshop, and P.~Nason for
interesting discussions during the workshop. This work was supported in part by
EU TMR  contract FMRX-CT98-0194 (DG 12 - MIHT).
\end{theacknowledgments}

% choose bibtex style depending on layout style and options used in
% sample:

\doingARLO[\bibliographystyle{aipproc}]
          {\ifthenelse{\equal{\AIPcitestyleselect}{num}}
             {\bibliographystyle{arlonum}}
             {\bibliographystyle{arlobib}}
          }
\bibliography{mf}

\begin{thebibliography}{24}
\expandafter\ifx\csname natexlab\endcsname\relax\def\natexlab#1{#1}\fi
\providecommand{\enquote}[1]{``#1''}
\expandafter\ifx\csname url\endcsname\relax
  \def\url#1{\texttt{#1}}\fi
\expandafter\ifx\csname urlprefix\endcsname\relax\def\urlprefix{URL }\fi

\bibitem[Forte(1999)]{threv}
Forte, S., \textbf{{\tt hep-ph/9910397}} (1999).

\bibitem[Chekelian(2001)]{exprev}
Chekelian, V., \textbf{{\tt hep-ph/0107053}} (2001).

\bibitem[Chekanov et~al.(2001)]{zeus}
Chekanov, S., et~al., \emph{{\rm ZEUS Coll.}}, \textbf{{\tt hep-ex/0105090}}
  (2001).

\bibitem[Adloff et~al.(2000)]{hone}
Adloff, C., et~al., \emph{{\rm H1 Coll.}}, \textbf{{\tt hep-ex/0012053}}
  (2000).

\bibitem[Ball and Forte(1994)]{das}
Ball, R.~D., and Forte, S., \emph{Phys. Lett.}, \textbf{B335}, 77--86 (1994).

\bibitem[Rujula et~al.(1974)]{DGPTWZ}
Rujula, A.~D., et~al., \emph{Phys. Rev.}, \textbf{D10}, 1649 (1974).

\bibitem[Forte and Ball(1995)]{zako}
Forte, S., and Ball, R.~D., \emph{Acta Phys. Polon.}, \textbf{B26}, 2097--2134
  (1995).

\bibitem[Mankiewicz et~al.(1997)]{lech}
Mankiewicz, L., Saalfeld, A., and Weigl, T., \emph{Phys. Lett.}, \textbf{B393},
  175--180 (1997).

\bibitem[Lipatov(1976)]{bfkla}
Lipatov, L.~N., \emph{Sov. J. Nucl. Phys.}, \textbf{23}, 338--345 (1976).

\bibitem[Fadin et~al.(1975)]{bfklb}
Fadin, V.~S., Kuraev, E.~A., and Lipatov, L.~N., \emph{Phys. Lett.},
  \textbf{B60}, 50--52 (1975).

\bibitem[Kuraev et~al.(1976)]{bfklc}
Kuraev, E.~A., Lipatov, L.~N., and Fadin, V.~S., \emph{Sov. Phys. JETP},
  \textbf{44}, 443--450 (1976).

\bibitem[Ball and Forte(1995)]{summing}
Ball, R.~D., and Forte, S., \emph{Phys. Lett.}, \textbf{B351}, 313--324 (1995).

\bibitem[Ellis et~al.(1995)]{EHW}
Ellis, R., Hautmann, F., and Webber, B., \emph{Phys. Lett.}, \textbf{B348},
  582--588 (1995).

\bibitem[Fadin and Lipatov(1998)]{fl}
Fadin, V.~S., and Lipatov, L.~N., \emph{Phys. Lett.}, \textbf{B429}, 127--134
  (1998).

\bibitem[Altarelli and Parisi(1977)]{ap}
Altarelli, G., and Parisi, G., \emph{Nucl. Phys.}, \textbf{B126}, 298 (1977).

\bibitem[Gribov and Lipatov(1972)]{gl}
Gribov, V.~N., and Lipatov, L.~N., \emph{Yad. Fiz.}, \textbf{15}, 781--807
  (1972).

\bibitem[McLerran(2001)]{lmcl}
McLerran, L., \textbf{{\tt hep-ph/0104285}} (2001).

\bibitem[Ball and Forte(1999)]{sxap}
Ball, R.~D., and Forte, S., \emph{Phys. Lett.}, \textbf{B465}, 271--281 (1999).

\bibitem[Altarelli et~al.(2000)]{sxres}
Altarelli, G., Ball, R.~D., and Forte, S., \emph{Nucl. Phys.}, \textbf{B575},
  313--329 (2000).

\bibitem[Ciafaloni et~al.(1999)]{ccs}
Ciafaloni, M., Colferai, D., and Salam, G.~P., \emph{Phys. Rev.}, \textbf{D60},
  114036 (1999).

\bibitem[Ball and Forte(1997)]{afp}
Ball, R.~D., and Forte, S., \emph{Phys. Lett.}, \textbf{B405}, 317--326 (1997).

\bibitem[Altarelli et~al.(2001{\natexlab{a}})]{sxphen}
Altarelli, G., Ball, R.~D., and Forte, S., \emph{Nucl. Phys.}, \textbf{B599},
  383--423 (2001{\natexlab{a}}).

\bibitem[Altarelli et~al.(2001{\natexlab{b}})]{sxrun}
Altarelli, G., Ball, R.~D., and Forte, S., \textbf{{\tt hep-ph/0109178}}
  (2001{\natexlab{b}}).

\bibitem[Salam(1998)]{salam}
Salam, G.~P., \emph{JHEP}, \textbf{07}, 019 (1998).

\end{thebibliography}

\end{document}